\documentstyle[aps,prb,epsf]{revtex}
\begin{document}
\twocolumn[\hsize\textwidth\columnwidth\hsize\csname
@twocolumnfalse\endcsname
\title{Temperature dependences of the surface resistance and the diamagnetic shielding susceptibility
 at $T_c-T<< T_c$ for high-$T_c$ superconductors$^{\ast\ast}$}
\author{Hyun-Tak Kim $^{\ast}$, Kwang-Yong Kang, and Seok-Kil
Han}
\address{Telecom. Basic Research Lab., ETRI, Taejon 305-350, Korea\\}
\maketitle{}
\begin{abstract}
At $T_c-T<< T_c$ (i.e., near $T_c$), in order to demonstrate the
conduction mechanism and temperature dependencies of the
diamagnetic-shielding susceptibility and the penetration depth, we
fabricated Ba$_{1-x}$K$_x$BiO$_3$ (BKBO) thin films and measured
the energy gap by tunnel effect and shielding susceptibilities
which are compared with those measured for BKBO and YBCO single
crystals. The shielding susceptibilities for BKBO and YBCO
crystals well-fit $\chi(T)/\chi(0)=1-exp(-2\triangle(T)/k_BT)$,
while that for the BKBO film follows $\chi(T)/\chi(0)=(1-T/T_c)$
which may not be intrinsic. The exponential decrease of the
susceptibilities near $T_c$ indicates that the conduction
mechanism is hopping. The energy gaps are observed as
${2{\triangle}(0)=(3.5{\pm}0.1) k_BT_c}$ for the BKBO film by the
tunnel effect, ${2{\triangle}(0)=(3.9{\pm}0.1) k_BT_c}$ for the
BKBO single crystal, and $2{\triangle}(0)=(8{\pm}0.2) k_BT_c$ for
the YBa$_{2}$Cu$_{3}$O$_{7}$ single crystal. Furthermore, for
microwave device applications of superconductors, at $T_c-T<<
T_c$, the surface resistance
$R_s(T){\approx}\sqrt{\frac{\omega\mu_0/2}{\sigma_n+(\sigma_s(0)-\sigma_n)f(T)}}$
 is derived from the surface impedance at $\omega\tau_{tr}<<$1,
 where $\sigma_s(0)$ and $\sigma_n$ are the
conductivities of the superconducting state and the normal state,
respectively, and
$f(T)=\chi(T)/\chi(0)=(1-exp(-2\triangle(T)/k_BT)$).
\\ \\ \\
\end{abstract}
]
\newpage
\section{INTRODUCTION}
The shielding susceptibility, $\chi$, was suggested as a means of
observing the London penetration depth, $\lambda_L$.$^{1-3}$
$\lambda_L$ has been known as a direct measure of the superfluid
density and a probe of the pairing state of superconductors. For
experimental results for high-$T_c$ superconductors at $T<<T_c$,
${\lambda}_L(T)$ has exhibited a $T^2$ temperature dependence$^4$,
as well as a linear $T$ dependence$^5$. The linear T dependence
has been predicted in $d$-wave superconductivity. Recently,
${\lambda}_L(T)$, which is independent of temperature, was
demonstrated.$^3$

On the other hand, for the 3 dimensional(D) XY behavior which
enhances fluctuations in the order parameter $\Psi$ for
Ginzburg-Landau (GL) theory, the diamagnetic shielding
susceptibility was given by
$\chi(T)=-(kT/{\Phi}_0^2)(T-T_c)^{-0.67}$, while the GL result
$\chi(T)\approx (T-T_{co})^{-0.5}$ near $T_c$.$^6$. The GL result
was suggested to need modification [6]. The 3D XY behavior was
absent in the magnetic penetration depth of
YBa$_2$Cu$_3$O$_{7-\delta}$ (YBCO) films.$^7$ Moreover, at
$T_c-T<< T_c$, from the BCS picture $\chi(T)/\chi(0)=(1-T/T_{c})$
was derived and from the phenomenological model
$\chi(T)/\chi(0)=1 - exp(-\frac{2\triangle(T)}{k_BT})$ was
obtained.$^{3,8,9}$

Furthermore, since 1960's, it has been known that BCS calculations
deviate from experimental data of the electronic specific-heat
contribution $C_{es}$ for superconductive lead above the
condensation temperature [10]. This suggests that a conduction
mechanism instead of tunneling one needs at $T_c-T<< T_c$.

For microwave device applications of superconductors, the
measurement and the physical analysis of the surface impedance are
very important.$^{11}$ In particular, its temperature dependence
is still controversial at $T_c-T<< T_c$,$^{12}$, since the
conduction mechanism with respect to conduction carriers has not
been clearly understood at $T_c-T<< T_c$.

In this paper, we demonstrate experimentally the unclear
temperature dependencies of $\chi(T)$ and ${\lambda}_L(T)$ at
$T_c-T<< T_c$ for shielding susceptibilities measured for a BKBO
film, high quality Ba$_{1-x}$K$_{x}$BiO$_3$ (BKBO) and YBCO
crystals. The energy gap of the BKBO film is obtained from the
tunnel effect, while energy gaps of BKBO and YBCO crystals are
obtained from the fitting parameter of shielding susceptibilities.
For microwave device applications of superconductors, the
temperature dependence of the surface resistance is derived.\\

\section{THEORETICAL CONSIDERATION}
The superconducting state as Meissner state is a condensed state.
With increasing temperature, Cooper pairs surpass the
superconducting energy gap to the excited state by the excitation
of thermal phonons; this means pair-breaking which characterizes
a superconductor-metal transition. In assuming that the
diamagnetic-shielding susceptibility is proportional to the
superfluid density in the superconducting
state~($\chi{\propto}n_{sup}$), in the microscopic theory
,$^{3,8}$ the susceptibility was given by,

at $T<<T_c$,
\begin{eqnarray}
{\chi(T)/\chi(0)} \approx 1 -
\sqrt{(2\pi\triangle(0)/k_BT)}e^{-\triangle(0)/k_BT} ,
\end{eqnarray}

and, at $T_c-T<< T_c$,
\begin{eqnarray}
\chi(T)/\chi(0)=(1-T/T_{c}).
\end{eqnarray}

Furthermore, at $T_c-T<<T_c$, another susceptibility was given by
\begin{eqnarray}
\chi(T)= \chi(0){(1 - exp (-\frac{2\triangle(T)}{k_BT}))}.
\end{eqnarray}

Equation (3) was obtained from the simple phenomenological model
based on the classical limit of Fermi-Dirac(FD) statistics
($i.e.$, Maxwell-Boltzman statistics) because the system is far
from condensation; in the system the fluid density is very small
and temperature is high relatively to $T_c$. Eq. (3) is the
hopping one of the conduction mechanism for semiconductors and was
given elsewhere.$^{3,9}$

The susceptibility as a function of the penetration depth has
been defined by
\begin{eqnarray}
{{\chi}(T)/{\chi}(0)=({\lambda}_L(0)/{\lambda}_L(T))^2}.
\end{eqnarray}\\
In addition, Eq (1) was demonstrated in a previous paper$^3$.

\begin{figure}
\vspace{0.1cm} \centerline{\epsfxsize=8.0cm\epsfbox{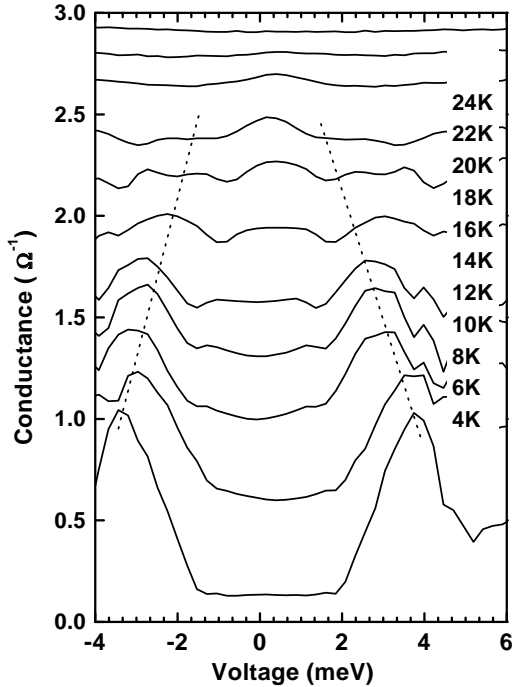}}
\vspace{0.1cm} \caption{Temperature dependence of the conductance
measured by the tunnel effect for the BKBO25 thin film; from 4 to
24 K.}
\end{figure}

\section{EXPERIMENT}
BKBO thin films were fabricated on (100)SrTiO$_3$ substrates in
the argon atmosphere of 1 Torr at the substrate temperature of
520$^{\circ}$ by laser ablation. The deposition method and
conditions were shown in a previous paper$^{13,14}$. The magnetic
shielding susceptibilities corresponding to the zero field cooled
(ZFC) susceptibility were measured at 10 Oe//c-axis by using
Quantum Design SQUID. The superconducting magnet of the SQUID was
quenched by perfect evaporation of liquid helium before the
measurement. Moreover the ZFC susceptibilies for BKBO and YBCO
single crystals were measured by the same method as that for the
BKBO film.

In order to measure the temperature dependence of the
superconducting energy gap for BKBO films, the conductances were
measured by the point-contact method through the
metal-superconductor-metal junction. The measurement temperatures
were from 4.2 K to 24 K.

\begin{figure}
\vspace{0.1cm} \centerline{\epsfxsize=8.5cm\epsfbox{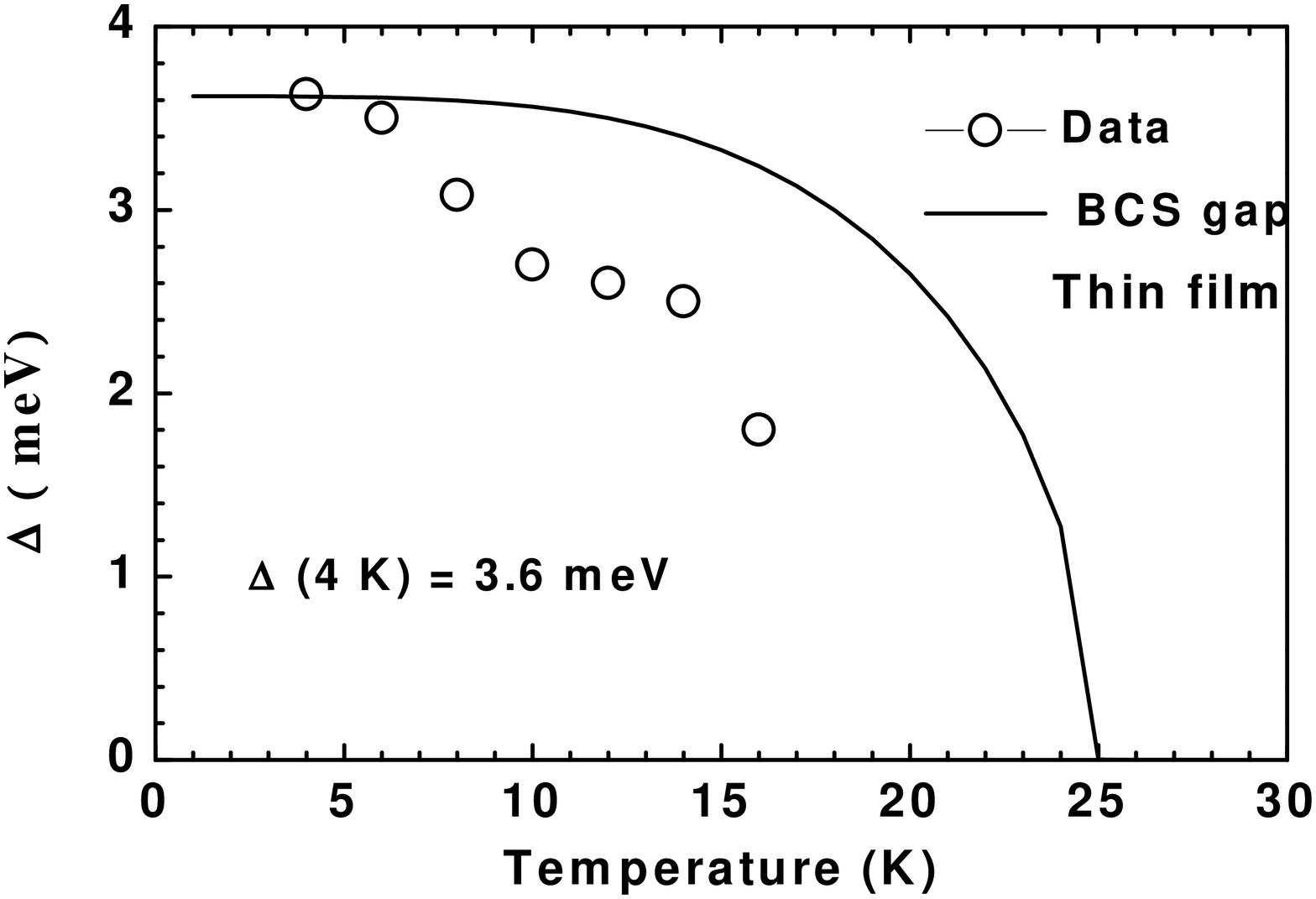}}
\vspace{0.1cm} \caption{Comparison of the energy gap measured for
the BKBO25 film with the empirical energy gap of the BCS one,
${\triangle}(T)/{\triangle}(0){\approx}
{\sqrt{cos[({\pi}/2)(T/T_c)^2]}}$ and $2{\triangle}(0)=bk_BT_c$.
Here $b{\approx}3.5$.}
\end{figure}

\section{RESULTS and DISCUSSION}
In order to check the validity of the linear T dependence in Eq.
(2), the temperature dependence of the conductance (or energy
gap), measured by the tunnel effect for a BKBO thin film (named
BKBO25) with $T_c\approx$25 K, is shown in Fig. 1. The
conductance at 4 K has a clean U shape which is known as the
energy gap with $\triangle$(4 K)$\approx$3.6 meV. U curve begins
to be deformed at 6 K, is observed up to 12 K, and is not seen
from 16 to 24 K. This indicates that the energy gap forms below
14 K. Fig. 2 shows the comparison of the measured energy gap with
the BCS gap with $b\approx3.5$. The measured gap deviates from
the BCS gap with increasing temperature. The deviation may arise
from the pinning effect or vortices due to impurity phases in the
film. Fig. 3 shows temperature dependencies of the susceptibility
obtained from both Eq. (3) and $\triangle$(T) determined from the
tunnel effect in Fig. 2, (called the "tunnel susceptibility"),
the susceptibility obtained from both Eq. (3) and the BCS gap,
(called the "BCS susceptibility"), and the ZFC susceptibility
data measured for the BKBO25 film at 10 Oe. Below 14 K, the
normalized susceptibilities agree closely. The figure shows that
the ZFC susceptibility data with linear T dependence at $T_c-T<<
T_c$ follow the tunnel susceptibility and deviate from the BCS
susceptibility. This indicates that crystals or films with an
energy gap which does not follow the BCS gap at $T_c-T<< T_c$
exhibit a linear $T$ dependence in the susceptibility, and that
Eq. (2) is not intrinsic.

\begin{figure}
\vspace{0.1cm} \centerline{\epsfxsize=8.5cm\epsfbox{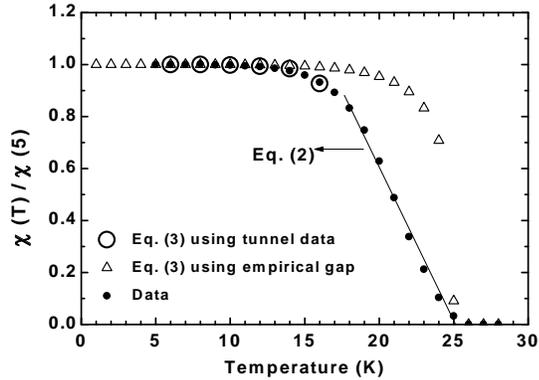}}
\vspace{0.1cm} \caption{Temperature dependencies of the
susceptibility (${\circ}$) obtained from Eq. (3) using the energy
gap determined from the tunnel effect in Fig. 1; the
susceptibility (${\triangle}$) obtained from Eq. (3) using the
empirical gap; and the ZFC susceptibility data (${\bullet}$)
measured directly at 10 Oe for the BKBO25 film. At $T_c-T<< T_c$,
the linear is Eq. (2).}
\end{figure}

At $T_c-T<< T_c$, the susceptibility data measured for the BKBO
single crystal are well fitted by Eq. (3) using the empirical gap
(${\triangle}(T)/{\triangle}(0){\approx}{\sqrt{cos[({\pi}/2)(T/T_c)^2]}}$)
of the BCS gap, as shown by the thick line in Fig. 4. The energy
gap is obtained from the fitting parameter
($2{\triangle}(0)=bk_BT_c$)as well and is given as
$2{\triangle}(0)=(3.9{\pm}0.1) k_BT_c$. The energy gap of the
single crystal agrees well with ${2{\triangle}(0){\approx}3.8
k_BT_c}$ which is determined by the tunnel effect [15]. Here,
although Eq. (3) is well-fitted from low temperatures to $T_c$,
the fitting cannot be believed below a temperature ($T{\approx}$17
K) in which the energy gap deviate from the BCS gap, as shown in
the inset of Fig. 4.

Figure 5 shows the susceptibility measured at 10 Oe//c-axis for a
YBCO single crystal like cube with the size of
$1.3{\times}1.1{\times}1~mm^3$ and $T_c{\approx}$93.2 K, grown by
a melting method. Eq. (1) and Eq. (3) are well applied to the
susceptibility data below 45 K and above 80 K, respectively, as
shown in Fig. 5 and its inset. The susceptibility in the
intermediate temperature range from 45 to 80 K is not explained by
the combined Eqs. (1) and (3), which is different from the BKBO
case. The 45 K is regarded as the condensation temperature.

\begin{figure}
\vspace{0.1cm} \centerline{\epsfxsize=8.5cm\epsfbox{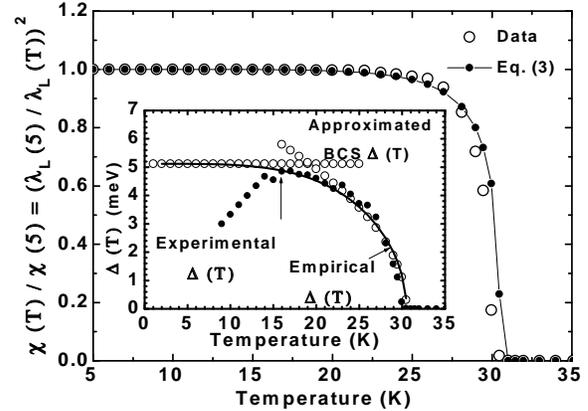}}
\vspace{0.5cm} \caption{The ZFC susceptibility measured at 10
Oe//c-axis for the BKBO single crystal. The open circles ($\circ$)
are experimental data. The thick line and solid circles
(${\bullet}$) indicates the fitting of Eq. (3) using the empirical
energy gap. In the inset, the approximated BCS gap is
${\triangle}(T)=3.2k_BT_c{\sqrt{1-T/T_c}}$ near $T_c$ and
$2{\triangle}(0)=bk_BT_c$ at $T$=0 K; the empirical gap is
${\triangle}(T)/{\triangle}(0){\approx}{\sqrt{cos[({\pi}/2)(T/T_c)^2]}}$;
the experimental gap is deduced from Eq. (3) and is
${\triangle}(T)= -(k_BT/2)ln(1-\chi(T)/\chi(0))$ where
$\chi(T)/\chi(0)$ uses the experimental susceptibility data. Here,
$b$=3.9.}
\end{figure}

\begin{figure}
\vspace{0.1cm} \centerline{\epsfxsize=8.5cm\epsfbox{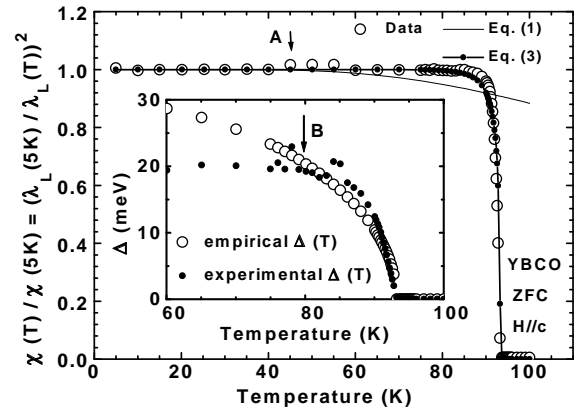}}
\vspace{0.5cm} \caption{Temperature dependence of the ZFC
susceptibility measured at 10 Oe//c-axis and its fitting for the
YBCO single crystal with  $2{\triangle}(0) {\approx}8 k_BT_c$ and
$T_c{\approx}$93.2 K. The susceptibility data (${\circ})$; Eq. (1)
(thin line); and Eq. (3) (thick line and solid circles
(${\bullet}$)). The inset shows the empirical gap and the
experimental one.}
\end{figure}

\section{SURFACE RESISTANCE}
The temperature dependence of the surface resistance, $R_s(T)$,
can be calculated by using the above results, which is necessary
for microwave device applications of superconductors. Eq. (3) was
derived by assuming $\chi(T)/\chi(0)=n_{sup}(T)/n_{tot}=f(T)$ and
${exp(-2\triangle(T)/k_BT)=n_{th}/n_{tot}}$,$^{3,8,9}$, where
$n_{tot}=n_{sup}+n_{th}$, $n_{tot}$ is the total number of pairs,
$n_{sup}$ denotes the number of Cooper pairs, and $n_{th}$ stands
for the number of thermally excited pairs ($i.e.$, carriers in the
normal state). Eq. (1) was also given by
$\chi(T)/\chi(0)=n_{sup}(T)/n_{tot}=f(T)$. The temperature
dependence of $\chi(T)$ corresponds to that of Cooper pairs. The
above consideration is similar to the two fluid model.

From the London equation, the conductivity of the superconducting
state is given by
$\sigma_{s}=\frac{(n_{sup}(\omega,T)/n_{tot})}{j{\mu}{\omega}{\lambda}^2(0)}$
at $0<{\omega}<{\omega}_s=2{\triangle}(0)/{\hbar}$, by using Eq.
(4). Here, $n_{sup}(\omega,T){\approx}n_{sup}(T)$, because Eq.
(3) is valid at $\omega<<\omega_s$. The temperature dependence of
$\sigma_s (T)$ is given by, at $T<<T_c$,
\begin{eqnarray}
{\sigma}_s(0)=\frac{1}{j{\mu_0}{\omega}{\lambda}^2(0)},
\end{eqnarray}
because of $n_{th}{\approx}$0 and
$f(T)=(n_{sup}(T)/n_{tot}){\approx}1$ in Eq. (1), while, at
$T_c-T<< T_c$,
\begin{eqnarray}
{\sigma}_s(T)=f(T){\sigma}_s(0)=\frac{f(T)}{j{\mu_0}{\omega}{\lambda}^2(0)},
\end{eqnarray}
where $f(T)=1-exp(-2\triangle(T)/k_BT)$. Moreover, because
$n_{th}$ is large at $T_c-T<< T_c$, the normal conductivity,
$\sigma_n$, occurring by $n_{th}$ cannot be ignored. Thus the
total conductivity is given by
\begin{eqnarray}
{\sigma}_T(T)&=&(\frac{n_{sup}}{n_{tot}}){\sigma}_s(0)+(\frac{n_{th}}{n_{tot}}){\sigma_n},
\noindent \\ &=&f(T){\sigma}_s(0)+(1-f(T)){\sigma_n},
\end{eqnarray}

where ${\sigma}_n={\sigma}_0/(1+j{\omega}{\tau}_{tr})$ [12].

In order to obtain the surface impedance
$Z_s=\sqrt{j\omega\mu_0/\sigma_T}$ for high quality
superconductors below the condensation temperature, $T_s$, $Z_s$
is important for microwave device application; $T_s{\approx}$17K
for BKBO and $T_s{\approx}$80K for YBCO, as shown in Figs. (4) and
(5). The total conductivity is approximated by
${\sigma}_T{\approx}{\sigma_s}$, since $n_{th}$ can be ignored
below $T_s$ because it is very small. The surface impedance is
given by, at $T<<T_c$,
\begin{eqnarray}
Z_s=\sqrt{j\omega\mu_0/\sigma_s}{\approx}j\mu_0\omega\lambda(0),
\end{eqnarray}

while, at $T_c-T<< T_c$,
\begin{eqnarray}
Z_s{\approx}\sqrt{j\omega\mu_0/(f(T){\sigma}_s(0)+(1-f(T)){\sigma_n})},
\end{eqnarray}
 by using Eq. 8. At $T<<T_c$, $Z_s$ is called the surface reactance $X_s$. At
$T_c-T<< T_c$, $Z_s$ can be expressed as $Z_s=R_s +jX_s$ and the
surface resistance is given by
\begin{eqnarray}
R_s(T){\approx}\sqrt{\frac{\omega\mu_0/2}{\sigma_n+(\sigma_s(0)-\sigma_n)f(T)}}.
\end{eqnarray}

This $R_s$ is obtained by calculations of Eq. 10 at
$\omega\tau_{tr}<<$1, increases with increasing temperature up to
$T_c$, and depends upon $n_{sup}({\propto}f(T)$).

Furthermore, at the intermediate temperature range in which
$\chi(T)$ does not be explained by Eqs. (1) and (3), the
theoretical approach for $Z_s$ is beyond scope of this paper. More
detailed discussion with other models will be done in a separate
paper.\\

\section{CONCLUSION}
In conclusion, at $T_c-T<< T_c$ (above the condensation
temperature), the shielding susceptibilities for BKBO and YBCO
crystals well-fit $\chi(T)/\chi(0)=(1-exp(-2\triangle(T)/k_BT))$
instead of $\chi(T)/\chi(0)=(1-T/ T_c)$. The exponential decrease
of the susceptibility indicates that the conduction mechanism is
hopping. The calculated surface resistance depends upon the
temperature dependence of the number of Cooper pairs at $T_c-T<<
T_c$. Furthermore, we contend that the temperature dependence of
the diamagnetic shielding susceptibility is that of Cooper
pairs;$\chi(T){\propto}n_{sup}(T)$.

\begin{center}
\noindent{\bf ACKNOWLEDGEMENTS}
\end{center}
I would like to acknowledge Prof. J. W. Hodby and Dr. W.
Schmidbauer for providing high quality BKBO crystals for Fig. 4.

\end{document}